\documentclass[conference]{IEEEtran}
\IEEEoverridecommandlockouts
\usepackage{cite}
\usepackage{amsmath,amssymb,amsfonts}
\usepackage{algorithmic}
\usepackage{graphicx}
\usepackage{textcomp}
\usepackage{xcolor}
\usepackage{units}
\usepackage{booktabs}
\usepackage{array}
\usepackage{comment}
\usepackage{threeparttable}

\def\BibTeX{{\rm B\kern-.05em{\sc i\kern-.025em b}\kern-.08em
    T\kern-.1667em\lower.7ex\hbox{E}\kern-.125emX}}
\begin{document}

\title{3GPP NR V2X Mode 2d: Analysis of Distributed Scheduling for Groupcast using ns-3 5G LENA Simulator}

\author{\IEEEauthorblockN{Thomas Fehrenbach\IEEEauthorrefmark{1}, Luis Omar Ortiz Abrego\IEEEauthorrefmark{1}, Cornelius Hellge\IEEEauthorrefmark{1}, Thomas Schierl\IEEEauthorrefmark{1} and Jörg Ott\IEEEauthorrefmark{2}}
	\IEEEauthorblockA{\IEEEauthorrefmark{1} Fraunhofer Heinrich Hertz Institute (HHI),  Berlin, Germany.}
    \IEEEauthorblockA{\IEEEauthorrefmark{2} Technical University of Munich,
Garching bei München, Germany}
    }

\maketitle

\begin{abstract}
Vehicle-to-everything (V2X) communication is a key technology for enabling intelligent transportation systems (ITS) that can improve road safety, traffic efficiency, and environmental sustainability. Among the various V2X applications, platooning is one of the most promising ones, as it allows a group of vehicles to travel closely together at high speeds, reducing fuel consumption and emissions. However, it poses significant challenges for wireless communication, such as high reliability and low latency. In this paper, we evaluate the benefits of group scheduling, also referred to as Mode 2d, which is based on a distributed and scheduled resource allocation scheme that allows the group of cars to select resources from a configured pool without network assistance. We evaluated the scheme through simulations, and the results show that this approach can meet the reliability, low latency, and data rate requirements for platooning.
\end{abstract}


\section{Introduction}
Vehicle-to-everything (V2X) communication has evolved significantly from its inception within LTE standards to address increasingly complex challenges in intelligent transportation systems. Initially introduced as the ProSe feature in LTE Release 12/13 for public safety applications, it was designed to enable direct device-to-device communication \cite{lien_enhanced_2016}. LTE Release 14 extended this capability, introducing the first C-V2X standard and enhancing support for high mobility and broadcast-only modes \cite{molina-masegosa_lte-v_2017}. Subsequent improvements in Release 15 and the introduction of the first 5G V2X standard in Release 16 have further refined these capabilities by adding support for unicast, groupcast, and aperiodic traffic, thus catering to a broader range of automotive and traffic efficiency applications \cite{naik_ieee_2019}\cite{garcia_tutorial_2021}. However, this evolution has revealed critical resource allocation challenges that traditional autonomous sensing methods struggle to address effectively in dense vehicular environments.

The 5G NR Mode 2d \cite{intel_corporation_r1-1814260_nodate} \cite{fraunhofer_hhi_r1-1812399_2018} represents a significant leap in addressing these challenges by introducing a distributed group scheduled resource allocation scheme that allows vehicles to autonomously select resources from a configured pool without network assistance. Despite Mode 2d being proposed during 3GPP V2X standardization, there remains a critical lack of public quantitative analysis examining the performance trade-offs between this approach and traditional sensing-based methods, particularly for group communication scenarios such as platooning where high reliability and low latency are paramount \cite{naik_ieee_2019}. This gap in understanding limits the ability to make informed deployment decisions for next-generation intelligent transportation systems.

This paper addresses this research gap through comprehensive performance evaluation using the ns-3 5G LENA module, providing three key contributions:
\begin{enumerate}
\item Detailed performance benchmarks comparing Mode 2d group scheduling with traditional sensing approaches across multiple traffic scenarios,
\item Analysis of critical coexistence scenarios examining mixed Mode 2/Mode 2d deployments, and
\item Analysis of cross-technology impact between Mode 2d groups and legacy sensing-based users, quantifying conditions under which Mode 2d provides net system benefit.
\end{enumerate}

\section{NR V2X Mode 1 and Mode 2 Overview}

In this section, we provide an overview of the NR V2X Modes 1, 2, and 2d highlighting the fundamental concepts and design principles that underpin each mode's functionality for Vehicle-to-Everything (V2X) communication.

\subsection{NR V2X Mode 1}

Mode 1, also known as ``scheduled resource allocation,'' is based on network-controlled resource scheduling, where a base station (gNB) manages resource assignments for V2X communications in response to user equipment (UE) requests. This centralized approach involves the following steps:

\begin{enumerate}
\item UE sends a resource request comprising its communication requirements and traffic demand.
\item Based on the received request, gNB computes an appropriate resource allocation for the UE.
\item The calculated assignment of resources is then relayed to the UE concerned through a control message.
\item Leveraging the allocated resources, the UE actively engages in V2X communication.
\end{enumerate}

Mode 1 is attractive due to its predictable and tightly controlled resource allocation, which can reduce interference, guarantee Quality of Service (QoS), and efficiently utilise available spectrum resources. However, it also has drawbacks, such as increased signalling overhead and latency \cite{garcia_tutorial_2021} \cite{ali_3gpp_2021}.

\subsection{NR V2X Mode 2}

Mode 2, or ``autonomous resource selection,'' empowers UEs with autonomous resource allocation capabilities, bypassing the need for centralized control by the gNB. This decentralization results in faster response times, reduced signaling overhead, and greater flexibility. Mode 2 operates by enabling UEs to:

\begin{enumerate}
\item Sense and evaluate the channel conditions to determine the available resources.
\item Identify suitable resources based on predefined criteria without any assistance from the gNB.
\item Engage in V2X communication using autonomously selected resources without waiting for gNB approval.
\end{enumerate}

Although Mode 2 offers a more flexible and scalable solution compared to Mode 1, it may suffer from increased collisions and interference due to the lack of centralized control. However, both Mode 1 and Mode 2 serve different purposes and deployment scenarios within the broader NR V2X communication framework \cite{ali_3gpp_2021} \cite{bazzi_design_2021}. During the standardization of NR V2X in Rel 16 several variations were proposed for mode 2\cite{gonzalezPerformanceEvaluationOptimal2023}:
\begin{itemize}
    \item \textbf{Mode 2a:} A V2X UEs autonomously select resources for V2X communication, similar to LTE V2X Mode 4.
    \item \textbf{Mode 2b:} A V2X UE assists other UEs in resource allocation by providing feedback to enable efficient resource usage.
    \item \textbf{Mode 2c:} A V2X UE uses one or more preconfigured sidelink (SL) patterns within a defined resource pool.
    \item \textbf{Mode 2d:} A V2X UE manages the allocation of SL resources for other UEs, acting in a role similar to a gNB in Mode 1, providing centralized resource coordination.
\end{itemize}

In the following, we will have a closer look at Mode 2d.
\subsection{NR V2X Mode 2d}

\begin{figure}[h]
\centering
\includegraphics[width=0.5\textwidth]{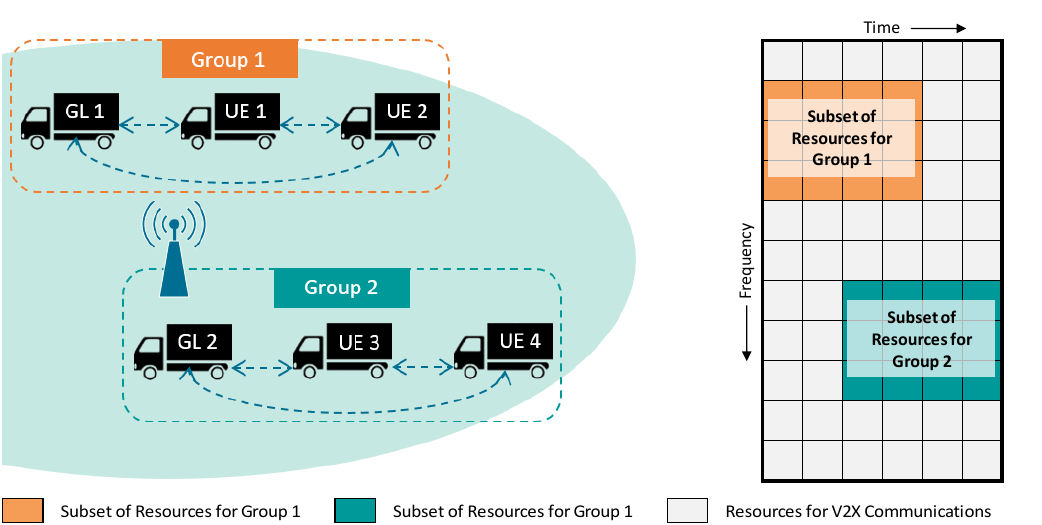}
\caption{Mode 2d.}
\label{fig:Mode2d}
\end{figure}
In Mode 2d\cite{fraunhofer_hhi_r1-1812399_2018} \cite{intel_corporation_r1-1814260_nodate}, as shown in Figure \ref{fig:Mode2d}, a group leader takes over the responsibility of resource management of a particular group. This reduces the reliance and burden on the gNB to allocate resources for every member of a group \cite{fraunhofer_hhi_r1-1812399_2018}.
This is specifically advantageous in situations where Mode 1 is not available or unreliable due to poor coverage or outages. In this case, the group needs to fallback to Mode 2 that traditionally relies on sensing for resource selection. 
With the proposed scheme, a group leader UE takes over in these cases to ensure efficient and reliable resource management.

\begin{figure}[h]
    \centering
    \includegraphics[width=0.5\textwidth]{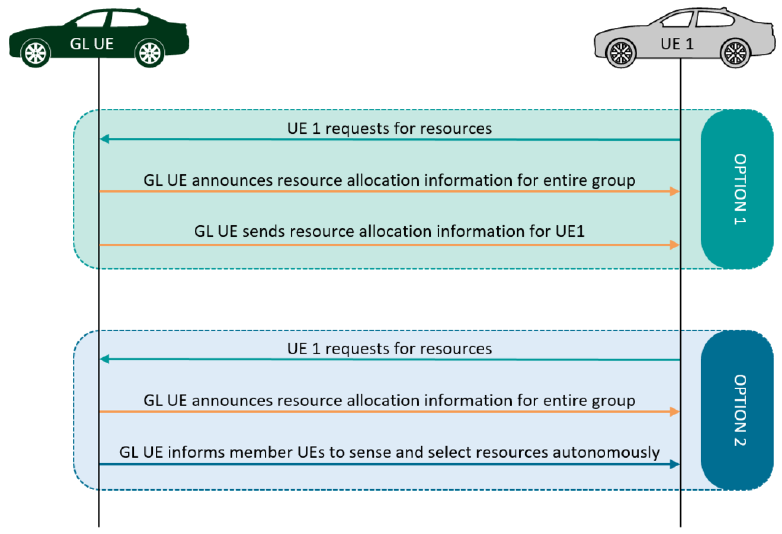}
    \caption{Mode 2d resource allocation options.}
    \label{fig:Mode2dRA}
    \end{figure}

There are different flavors of Mode 2d introduced in \cite{fraunhofer_hhi_r1-1812399_2018}. These can be seen in Figure \ref{fig:Mode2dRA}:
\begin{itemize}
    \item Option 1: The GL UE allocates the resources to be used by the individual member UEs. The reliability of these resources is very high since they are individually allocated to each member UE.
    \item Option 2: The GL UE informs the member UEs that they can carry out sensing within the set of selected resources for the group. Since only members of the group compete for these selected resources, the probability of collisions is low, and it avoids additional signalling overhead within the group.
\end{itemize}

\section{Platooning}
\subsection{Platooning Overview}
Platooning refers to a group of vehicles traveling closely together, each of which follows the one in front.  This study concentrates on the platooning scenario, which is a promising application for NR V2X Mode 2d. Within platooning, the vehicles within the cluster interact to maintain a safe gap and synchronize their acceleration and deceleration. In addition, the vehicles in the platoon communicate with the gNB to receive details about the upcoming road conditions and relay their speeds and locations \cite{SAEJ3016} \cite{ETSI_TS_122_186}.
\subsection{Platooning Requirements}
The 5G Automotive Association (5GAA) highlights Vehicle-to-Everything (V2X) platooning as a critical aspect of automotive communication systems, emphasizing the importance of meeting specific latency and reliability criteria. Platooning involves the synchronized movement of multiple interconnected vehicles in close proximity with minimal human involvement. Effective platooning requires seamless communication between vehicles and infrastructure, with a focus on achieving low latency and high reliability\cite{5gaa_5gaa_2021}. 

Table \ref{tab:communication_requirements} summarizes the performance requirements for different platooning scenarios. These requirements include payload size, transmission rate, maximum end-to-end latency, reliability, and minimum required communication range. The table highlights the stringent requirements necessary to ensure the safe and efficient operation of platooning systems \cite{ETSI_TS_122_186}.

\begin{table}[h]
  \centering
  \caption{Platooning Scenarios and Communication Requirements\cite{ETSI_TS_122_186}.}
  \label{tab:communication_requirements}
\begin{threeparttable}
\tiny 
  \begin{tabular}{|p{1.9cm}|p{0.7cm}|p{0.4cm}|p{.95cm}|p{0.62cm}|p{.59cm}|p{.45cm}|}
    \hline
    \textbf{Scenario} & \textbf{Automation} & \textbf{Payload (Bytes)} & \textbf{Tx Rate (Messages/s)} & \textbf{E2E Latency} & \textbf{Reliability} & \textbf{MCR} \\ \hline
    Cooperative driving for vehicle platooning & Lowest & 300-400 & 30 & 25 ms & 90\% & -\tnote{a} \\ \hline
    Information exchange between a group of UEs & Highest & 50-1200 & 30 & 10 ms & 99.99\% & 80 m \\ \hline
    Reporting for platooning (UEs and RSU) & N/A & 50-1200 & 2 & 500 ms & -\tnote{a} & -\tnote{a} \\ \hline
    Information sharing for platooning (UEs and RSU) & Lower & 6000 & 50 & 20 ms & -\tnote{a} & 350 m \\ \hline
  \end{tabular}
\begin{tablenotes}
\item[a] not specified.
\end{tablenotes}
\end{threeparttable}
  
\end{table}

Adherence to these requirements is crucial to improving traffic flow, reducing congestion, and improving road safety through V2X platooning, thus decreasing the likelihood of human errors while driving.
\begin{figure}
    \centering
    \includegraphics[width=1\linewidth]{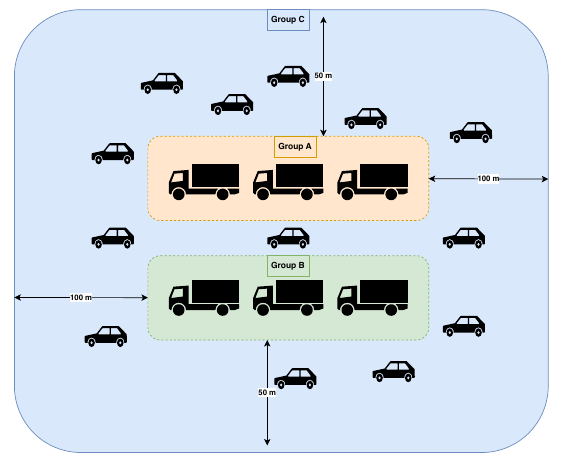}
    \caption{Simulation Setup.}
    \label{fig:groupABClayout}
\end{figure}

\begin{figure}
    \centering
    \includegraphics[width=1\linewidth]{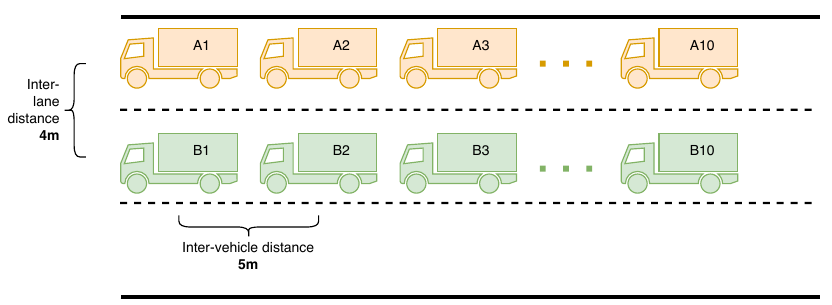}
    \caption{Platoon Setup.}
    \label{fig:highwayAB}
\end{figure}

\section{Simulation Setup}

To evaluate the efficacy of Mode 2d in a V2X environment, we used the ns-3 simulator with the 5G-LENA module \cite{ali_3gpp_2021}, configured to closely mimic the conditions of real-world vehicular networks (see Fig. \ref{fig:Mode2d} for vehicle arrangement). The simulation setup involved the following parameters:

\begin{itemize}
\item \textbf{Vehicle Grouping}: In our scenario vehicles are grouped into two main categories for groupcast communications, Group A and Group B, with a third category, Group C, handling background traffic.  This is visualized in Fig. \ref{fig:groupABClayout}.
\item \textbf{Communication Types}: Groupcast A and B were used to simulate typical V2X communications, while background traffic was simulated through Group C to assess the system's performance under varying traffic densities.
\item \textbf{Platooning}: Within each group, the vehicles are spaced at a uniform inter-vehicle distance of 5 meters between their centers. Furthermore, there is an inter-lane distance of 4 meters between the centers of vehicles in different groups. This is depicted in Fig. \ref{fig:highwayAB}
\item \textbf{Simulation Parameters}: Key Parameters are shown in Table \ref{table:simulation_parameters}.
\end{itemize}

It is important to note that this study focuses on pre-established, stable platooning groups without modeling group formation, merging, or splitting dynamics.

\subsection{Maximum Reuse Distance Scheduler for Mode 2d Group Scheduling}

The Mode 2d group scheduling, as implemented in our paper, utilizes a Maximum Reuse Distance (MRD) scheduler\cite{cecchini_maximum_2018} to optimize resource allocation in vehicular groups. It maximizes spatial reuse of the spectrum and minimizes resource collisions. This is crucial in dense vehicular networks for high communication reliability and low latency. The MRD scheduler allocates resources to maximize the distance between any two vehicles using the same resource within a group, focusing only on intra-group optimization. The MRD scheduler's operation is mathematically modeled as follows:

\begin{enumerate}
    \item \textbf{Resource Pool Configuration}: Let $\mathcal{R}$ denote the set of resources within the configured pool. 

    \item \textbf{Distance Calculation}: For a group of $m$ vehicles, let $d_{ij}$ represent the Euclidean distance between vehicle $i$ and vehicle $j$ within the same group, where $i, j \in \{1, 2, \dots, m\}$ and $i \neq j$. The distance $d_{ij}$ is calculated using the coordinates $(x_i, y_i)$ and $(x_j, y_j)$ of vehicles $i$ and $j$, respectively:

    \begin{equation}
    d_{ij} = \sqrt{(x_i - x_j)^2 + (y_i - y_j)^2}
    \end{equation}

    \item \textbf{Resource Allocation}: The MRD scheduler allocates resources iteratively whenever a group member requires a grant. When a vehicle $i$ requests a resource, the scheduler maximizes the minimum distance between two vehicles within the group sharing the same resource. $\mathcal{U}_r$ are the vehicles using a resource $r$. For each grant, the scheduler determines the resource by:

    \begin{equation}
    \arg\max_{r \in \mathcal{R}} \left( \min_{j \in \mathcal{U}_r} d_{ij} \right) 
    \end{equation}

    With some abuse of notation $\min_{j \in \mathcal{U}_r} d_{ij}$ is defined as $\infty$, if $\mathcal{U}_r = \emptyset$. This ensures that vehicles $i$ and $j$, which are assigned the same resource $r_k$, are separated by the maximum possible distance within the group.

    \item \textbf{Periodic Reselection}: In line with NR V2X principles, periodic reselection is incorporated to maintain optimal resource allocation over time, accommodating the dynamic nature of vehicular environments. Each resource $r_k$ is associated with a reselection period $T_{r_k}$, which defines the interval after which the allocation of the resource is re-evaluated. After $T_{r_k}$ expires, the vehicle currently using resource $r_k$ reassesses whether to continue using the same resource or switch to another available resource within the pool $\mathcal{R}$.
    
    The decision to reselect is governed by the reselection probability $P_{reselect}$, which reflects the probability that a resource change will occur, else the resource will be reused.
    This ensures that resources are periodically reassessed and reallocated to optimize the communication reliability and efficiency within the vehicular group.

\end{enumerate}

\subsection{Scenarios}

To evaluate the impact of different resource allocation strategies, three distinct scenarios were analyzed:

\begin{enumerate}
    \item \textbf{Scenario 1: All Groups Using Mode 2 (Sensing)}\\
    In this scenario, all vehicle groups rely on Mode 2, where each vehicle autonomously selects resources based on sensing. This setup simulates a fully decentralized environment without any centralized or group-based resource management, allowing us to observe the performance of traditional sensing methods under varying data rates.

    \item \textbf{Scenario 2: All Groups Using Mode 2d (Group Scheduling)}\\
    Here, all vehicle groups employ Mode 2d, utilizing a group leader to manage resource allocation within the group. This scenario examines the effectiveness of centralized, group-based resource scheduling in enhancing communication reliability and reducing interference compared to purely sensing-based methods.

    \item \textbf{Scenario 3: Mixed Scenario (One Group Using Mode 2 and the Other Mode 2d)}\\
    In this mixed scenario, one vehicle group operates under Mode 2 (sensing), while the other uses Mode 2d (group scheduling). This setup allows us to evaluate how the coexistence of different resource allocation strategies affects overall network performance, particularly focusing on the interaction between the scheduled and sensing groups.
\end{enumerate}

\begin{table}
\centering
\caption{Simulation parameters.}
\label{table:simulation_parameters}
\begin{tabular}{|>{\centering\arraybackslash}p{0.23\linewidth}|c|p{0.3\linewidth}|}
\hline
\textbf{Parameter} & \textbf{Value} & \textbf{Comments} \\ \hline
Numerology                      & 0 (15kHz)                    & Numerology type identified by the subcarrier spacing \\ \hline
Simulation Time                 & 60 s                         & Total simulation time in seconds \\ \hline
slTxTransNum                    & 3                            & Number of sidelink transmissions \\ \hline
MCS                             & 14                           & Modulation and coding scheme index \\ \hline
t1                              & 2                            & Time parameter t1 for scheduling \\ \hline
t2                              & 32                           & Time parameter t2 for scheduling \\ \hline
Packet Size                     & 300 byte                     & Size of the packet in bytes \\ \hline
Data rate & \begin{tabular}[c]{@{}c@{}}A/B 24 kbps\\ C 48 kbps\end{tabular} & Data rate for different categories A/B/C \\ \hline
Reservation Period              & 50 ms                        & Duration of reservation period for resources \\ \hline
Frequency                       & 5.9 GHz                      & Central frequency for the transmission \\ \hline
Bandwidth                       & 40 MHz                       & Bandwidth allocated for the signal \\ \hline
Tx Power                        & 23 dBm                       & Transmission power in dBm \\ \hline
Seeds/Simulations               & 100                          & Number of seeds for simulations to ensure statistical validity \\ \hline
\begin{tabular}[c]{@{}c@{}}Random Vehicles\\ (Group C)\end{tabular} & 1 to 170                     & Range of random vehicles in group C \\ \hline
Speed                           & 0 m/s                        & Relative speed of the transmitter and receiver \\ \hline
\end{tabular}
\end{table}

\section{Results}

Our simulation results provide a comprehensive assessment of Mode 2d's distributed scheduling on V2X communication performance across various traffic scenarios. This section delves into the key performance indicators. Packet Reception Rate (PRR) and Packet Inter-Reception Time (PIR) are evaluated under varied resource allocation strategies.

\subsection{Packet Reception Rate (PRR)}
Packet Reception Rate (PRR) measures the proportion of packets that are successfully received out of the total transmitted. This metric functions to evaluate the reliability and efficiency of the communication link, where higher PRR values reflect better performance. The PRR can be mathematically represented as:

\begin{equation}
\text{PRR} = \frac{N_{\text{received}}}{N_{\text{sent}}}
\end{equation}

where $N_{\text{received}}$ is the number of successfully received packets and $N_{\text{sent}}$ is the total number of packets transmitted.

\begin{figure}
    \centering
    \includegraphics[width=1\linewidth]{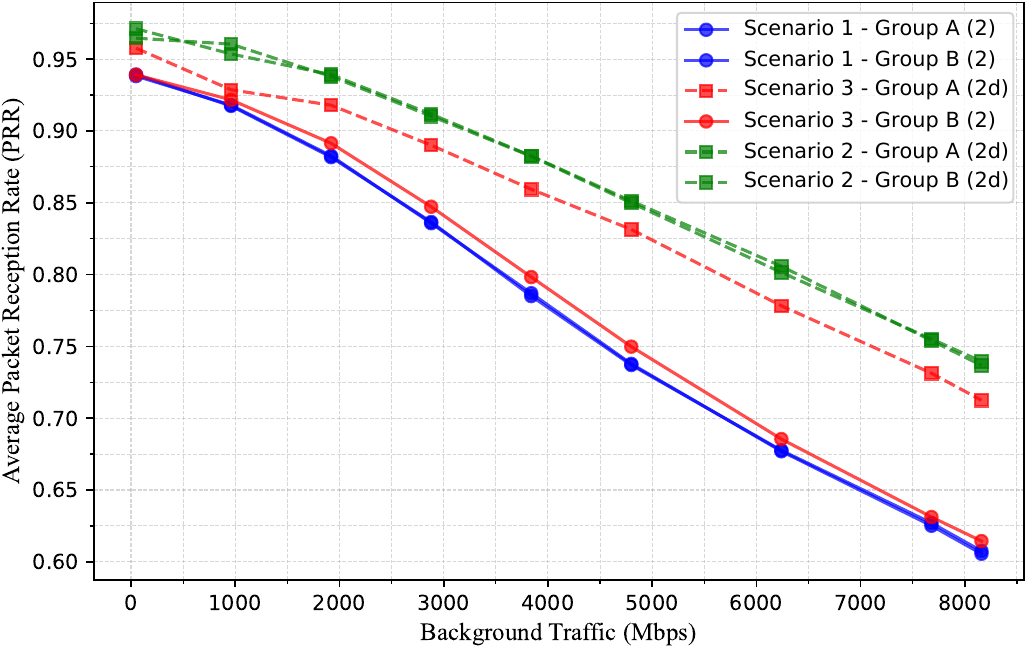}
    \caption{Comparison of groups A (solid) and B (dashed) at Scenarios 1 (blue), 2 (green) and 3 (red): (\textbf{2}) stands for sensing Mode 2 and (\textbf{2d}) for group scheduling Mode 2d.}
    \label{fig:PRR_all}
\end{figure}

\begin{figure}
    \centering
    \includegraphics[width=1\linewidth]{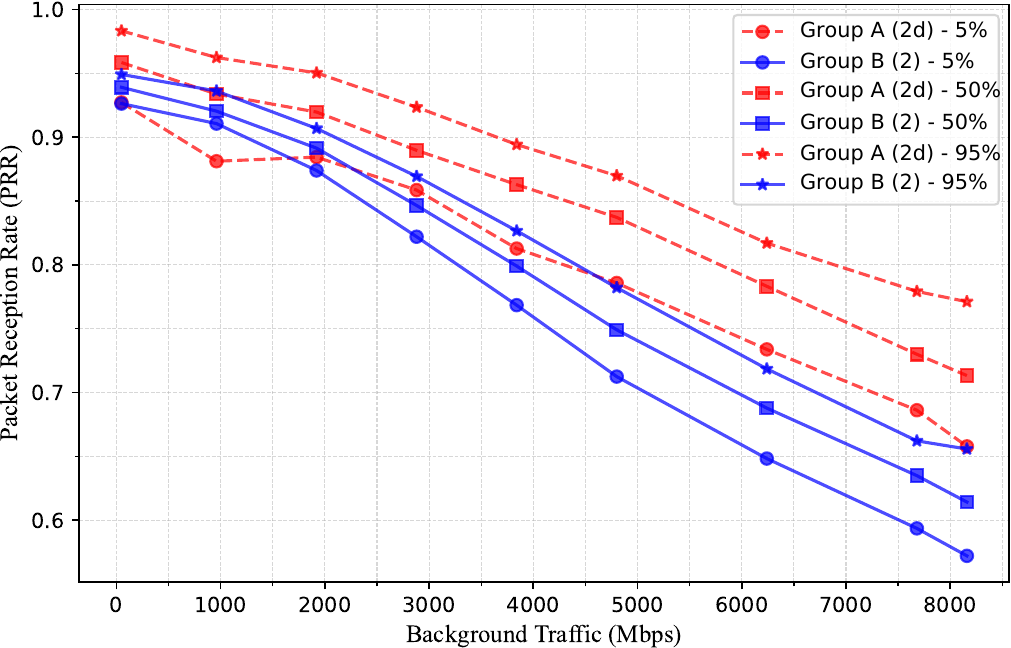}
    \caption{Performance analysis of Scenario 3 across percentiles: Group A (red) is group scheduling Mode 2d and Group B (blue) is sensing Mode 2.}
    \label{fig:PRR_3_Percentiles}
\end{figure}

\begin{figure}
    \centering
    \includegraphics[width=1\linewidth]{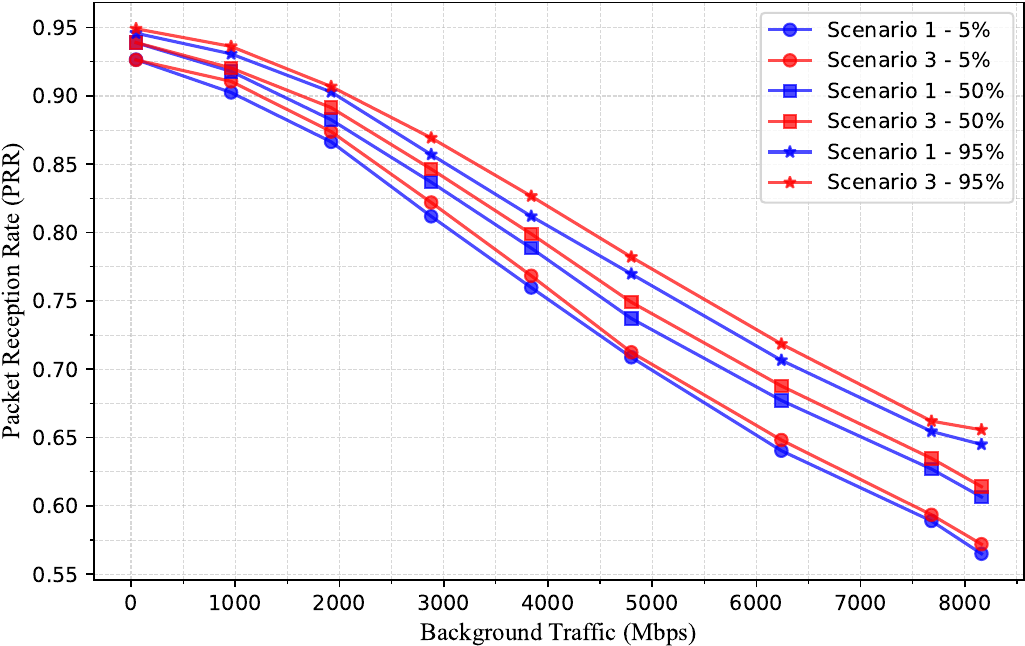}
    \caption{Percentile-based outcomes for Mode 2 between Scenarios 1 (blue) \& 3 (red).}
    \label{fig:PRR-1_3-Sensing_Percentile}
\end{figure}

PRR analysis reveals a significant advantage of Mode 2d scheduling over sensing-based and random resource selection methods. As depicted in Figures \ref{fig:PRR_all} our empirical data show that:
\begin{itemize}
    \item Group Scheduled scenarios consistently achieved the highest PRR, demonstrating Mode 2d's efficiency in orchestrating resources within specific vehicle groups and mitigating external interference.
    \item Mixed mode scenarios, which combined Mode 2d scheduling with sensing methods, exhibited enhanced PRR over scenarios based solely on sensing.
\end{itemize}

\subsection{Packet Inter-Reception Time (PIR)}
Packet Inter-Reception Time (PIR) represents the mean time gap between successive packets that are successfully received. It offers an understanding of the communication regularity and timing, where smaller PIR values imply more regular and frequent packet reception. The PIR is mathematically represented as:

\begin{equation}
\text{PIR} = \frac{1}{N_{\text{received}} - 1} \sum_{i=2}^{N_{\text{received}}} (t_{i} - t_{i-1})
\end{equation}

where $t_{i}$ and $t_{i-1}$ are the reception times of the $i$-th and $(i-1)$-th successfully received packets, respectively, and $N_{\text{received}}$ is the total number of packets received.

\begin{figure}
    \centering
    \includegraphics[width=1\linewidth]{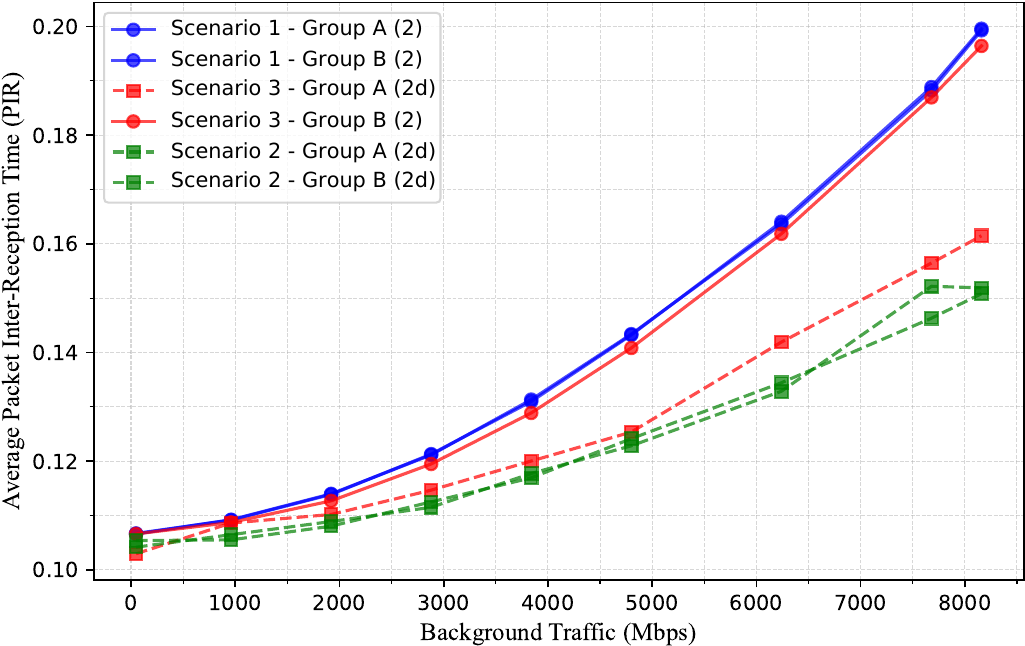}
    \caption{Comparison of groups A (solid) and B (dashed) withing the Scenarios 1 (blue), 2 (green) and 3 (red): (\textbf{2}) stands for sensing Mode 2 and (\textbf{2d}) for group scheduling Mode 2d.}
    \label{fig:PIR_All}
\end{figure}

\begin{figure}
    \centering
    \includegraphics[width=1\linewidth]{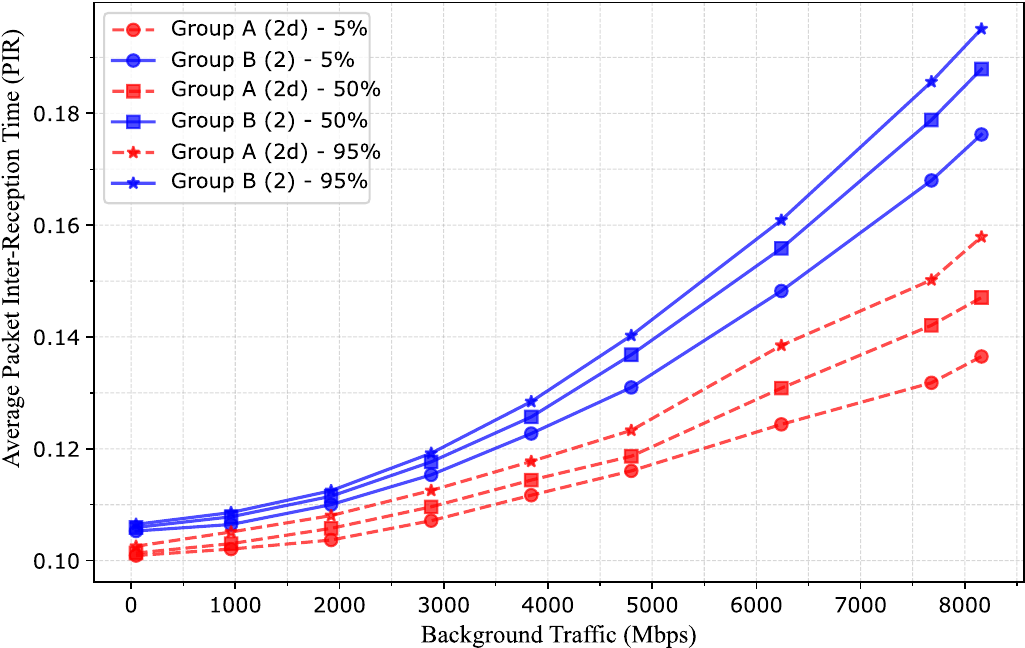}
    \caption{Performance analysis of both groups of Scenario 3 in terms of Average PIR across percentile thresholds: Group A (red) is group scheduling Mode 2d and Group B (blue) is sensing Mode 2.}
    \label{fig:PIR_3_Percentiles}
\end{figure}

\begin{figure}
    \centering
    \includegraphics[width=1\linewidth]{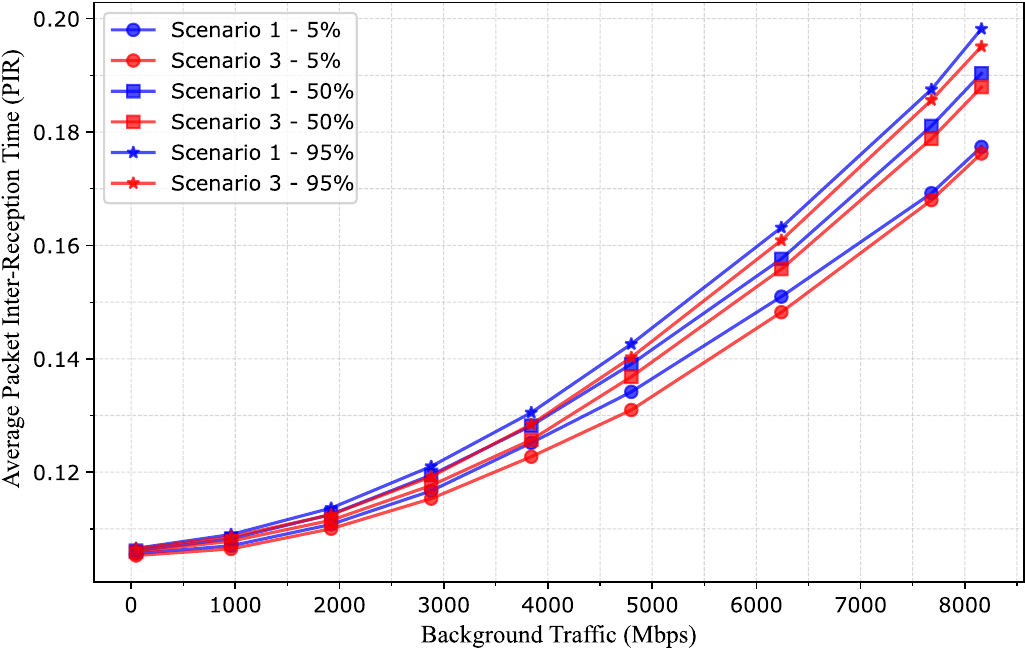}
    \caption{Percentile-based outcomes for Sensing groups in a Scenario 1 (blue) \& 3 (red).}
    \label{fig:PIR-1_3-Sensing_Percentiles}
\end{figure}

Our analysis of PIR further confirms the effectiveness of Mode 2d. As shown in Figures \ref{fig:PIR_All}:
\begin{itemize}
    \item Scheduled scenarios maintained lower PIR across varying traffic densities, indicating more consistent and dependable communication timings compared to other approaches.
    \item Mixed scenarios displayed variability in PIR, yet outperformed purely sensing-based methods, especially in high vehicle density contexts. This suggests that Mode 2d can effectively alleviate intermittent communication lags commonly observed in autonomous sensing arrangements.
\end{itemize}

\subsection{Impact of Mode 2d on Mode 2 Performance}

The simulation results demonstrate a significant interaction between Mode 2 and Mode 2d groups, where Mode 2 groups consistently benefit from the presence of Mode 2d groups across all UEs and percentile ranges. In Scenario 3 (Figure \ref{fig:PRR_3_Percentiles}), the scheduled group (Mode 2d) maintains significantly higher PRR values compared to the sensing group (Mode 2). Notably, the 5th percentile of PRR for scheduled UEs surpasses the 95th percentile of PRR for sensing UEs for high traffic densities, highlighting the superior reliability of Mode 2d's scheduling approach.

PIR analysis (Figure \ref{fig:PIR_3_Percentiles}) further supports these findings, showing that scheduled groups achieve lower inter-reception times and exhibit greater consistency, even under increasing traffic loads. The scheduled groups consistently maintain lower PIR values across all percentiles, indicating more reliable and frequent packet reception compared to sensing groups.

Figures \ref{fig:PRR-1_3-Sensing_Percentile} and \ref{fig:PIR-1_3-Sensing_Percentiles} demonstrate that Mode 2 sensing UEs benefit significantly from the presence of Mode 2d scheduled UEs across all traffic densities and percentile levels.

Mode 2d outperforms traditional sensing methods due to its centralized group scheduling, where a group leader (GL) allocates resources, significantly reducing collisions and optimizing resource efficiency in dense networks. The Maximum Reuse Distance (MRD) scheduler further enhances reliability by maximizing spatial reuse and minimizing resource conflicts. Additionally, periodic reselection dynamically adapts resource allocation to network changes, maintaining consistent communication quality. 

In mixed scenarios, Mode 2d confines resource usage within controlled subsets, reducing interference and improving the performance of coexisting Mode 2 groups. This structured allocation achieves higher Packet Reception Rates (PRR), lowers Packet Inter-Reception Time (PIR) variability, and supports sparse network deployments by offloading resource management from the gNB. 

These analyses collectively underscore the critical role of Mode 2d in enhancing V2X communication performance. The observed differences in PRR and PIR between sensing and scheduled groups confirm the advantages of Mode 2d's distributed scheduling mechanism over traditional autonomous sensing, positioning it as a pivotal enabler for next-generation vehicular networks.

\section{Conclusion}

This paper has presented an analysis of the NR V2X Mode 2d resource allocation mechanism using the ns-3 5G LENA simulator, with a focus on its application to platooning scenarios. The results demonstrate that Mode 2d significantly outperforms traditional sensing-based approaches in the key performance metrics Packet Reception Rate (PRR) and Packet Inter-Reception Time (PIR). The distributed and scheduled resource allocation of Mode 2d provides a robust solution to maintain high reliability and low latency communication in high-density vehicular networks, which are critical for the effective implementation of platooning and other advanced V2X applications.

Empirical data shows that the integration of Mode 2d into existing V2X frameworks can lead to a marked improvement in communication performance, even under mixed mode scenarios where traditional sensing mechanisms coexist. This highlights the potential for Mode 2d to be a cornerstone in the development of future vehicular communication systems, particularly in environments that require stringent communication requirements.

In conclusion, the adoption of Mode 2d within the 5G NR V2X standard presents a viable way forward to address the communication challenges posed by next-generation intelligent transportation systems. The demonstrated benefits suggest that Mode 2d should be considered a key component in the ongoing evolution of V2X technologies, offering a scalable and efficient solution to improve vehicle communication reliability and performance.

\bibliographystyle{IEEEtran}
\bibliography{references}

\end{document}